\documentclass[a4paper,11pt]{article}
\usepackage{jinstpub}   

\usepackage{graphicx}
\usepackage{booktabs}
\usepackage{multirow}
\usepackage{amsmath}
\usepackage{array}
\usepackage{longtable}
\usepackage{float}
\usepackage{xcolor}
\usepackage{tabularx}
\usepackage{enumitem}
\usepackage{listings}
\usepackage{microtype}
\usepackage{siunitx}
\usepackage{tcolorbox}
\usepackage{lineno}
\tcbuselibrary{breakable, skins}

\numberwithin{equation}{section}

\definecolor{ragdark}{HTML}{1F335C}
\definecolor{ragblue}{HTML}{4A86B8}
\definecolor{raggreen}{HTML}{2A9D8F}

\lstdefinestyle{code}{basicstyle=\ttfamily\small,breaklines=true,columns=fullflexible,frame=single,backgroundcolor=\color{gray!5},xleftmargin=0.5em,xrightmargin=0.5em}
\setlist[itemize]{leftmargin=1.5em}
\setlist[enumerate]{leftmargin=1.5em}

\title{Agentic hybrid RAG for evidence-grounded muon collider analysis}

\author[a]{Ruobing Jiang,}
\author[a,1]{Dawei Fu,\note{Corresponding author.}}
\author[b]{Cheng Jiang,}
\author[a]{Tianyi Yang,}
\author[a]{Zijian Wang,}
\author[a]{Youpeng Wu,}
\author[a]{Yong Ban,}
\author[a]{Yajun Mao}
\author[a]{and Qiang Li}

\affiliation[a]{State Key Laboratory of Nuclear Physics and Technology, Peking University,\\
  Beijing 100871, China}
\affiliation[b]{School of Physics and Astronomy, University of Edinburgh,\\
  Edinburgh EH9 3FD, United Kingdom}

\emailAdd{fudw@pku.edu.cn}

\abstract{Muon collider research spans accelerator physics, detector instrumentation, and high-energy phenomenology, with relevant evidence scattered across a rapidly expanding and heterogeneous body of scientific literature. As high-energy physics (HEP) increasingly explores agent-assisted analysis workflows, efficiently locating, integrating, and verifying scientific evidence becomes an essential capability. While retrieval-augmented generation (RAG) offers a promising framework for scientific question answering, integrating agentic reasoning without compromising retrieval precision remains a key challenge. In this work, we present agentic hybrid RAG, an evidence-grounded RAG framework for muon collider research. The framework combines a hybrid retriever, integrating sparse lexical and dense semantic retrieval, with an agentic reasoning module for query decomposition, evidence expansion, and grounded answer generation. To enable systematic evaluation, we construct the first benchmark for retrieval-augmented scientific question answering in the muon collider domain, comprising a curated literature corpus together with dedicated retrieval and answer-generation benchmarks covering major detector and physics research topics. Extensive evaluation shows that hybrid retrieval provides the strongest retrieval backbone, while agentic reasoning is most effective for controlled evidence expansion and answer synthesis. Built on this principle, agentic hybrid RAG consistently outperforms representative retrieval and RAG baselines in retrieval effectiveness, answer quality, evidence coverage, and factual grounding. Together, the benchmark and framework provide a foundation for evidence-grounded scientific question answering and future HEP analysis agents operating over large-scale scientific literature. The code and benchmarks are publicly available at
\url{https://github.com/AItutorialjrb/RAG_muon_JINST}.

}


\arxivnumber{2606.10381}

\begin{document}
\maketitle
\flushbottom

\section{Introduction}
As AI agents become increasingly explored for HEP research and analysis workflows~\cite{gendreaudistler2025automatinghighenergyphysics,moreno2026aiagentsautonomouslyperform}, there is a growing need for systems that can reliably access, retrieve, and utilize scientific literature. Analysis decisions rarely depend on isolated facts but instead on evidence distributed across a large and rapidly evolving body of publications, making literature retrieval and interpretation a central capability of next-generation HEP analysis agents.

Building such agents is non-trivial: they must support reliable evidence extraction and synthesis over long-tail and rapidly evolving knowledge, where relevant information is often fragmented across multiple papers and subfields. Large language models (LLMs) have been widely explored for this purpose but remain limited in producing faithful, evidence-grounded outputs without explicit external grounding~\cite{ji2023survey,lewis2020rag}. Retrieval-augmented generation (RAG) addresses this by grounding generation in external corpora, reducing reliance on parametric memory and costly model retraining~\cite{lewis2020rag}.

Effective retrieval for HEP literature requires integrating complementary signals. Dense semantic retrieval can effectively capture semantic similarity and match paraphrased queries~\cite{karpukhin2020dpr, reimers2019sentencebert}, but it may overlook exact acronyms, mathematical symbols, and process names that are prevalent in HEP. Sparse lexical retrieval methods, exemplified by Okapi Best Matching 25 (BM25)~\cite{robertson-2009}, provide robust keyword-level matching and are particularly effective for terminology-sensitive queries. Hybrid strategies combining these signals via reciprocal rank fusion (RRF)~\cite{cormack2009reciprocal} consistently improve robustness across diverse query types. Recent agentic RAG extensions incorporate query decomposition and routing~\cite{singh2025agenticrag}, but excessive agentic exploration can introduce retrieval drift, motivating architectures that balance hybrid retrieval with lightweight agentic reasoning.

Muon collider research provides a concrete and challenging testbed
for such systems. It spans accelerator physics, detector instrumentation, and high-energy phenomenology, drawing on beam dynamics, machine-detector interfaces (MDI), beam-induced backgrounds (BIB), detector design, and physics analyses. The field is advancing rapidly: following renewed international interest since the early 2020s, the volume of technical reports, conference proceedings, and preprints has grown substantially, with contributions distributed across accelerator, detector, and phenomenology communities~\citep{muoncollider_pbc2020, imcc2022}. Information relevant to a single question is often distributed across multiple papers and subfields: understanding a detector-design choice, for instance, may require connecting evidence from background studies, shielding reports, and performance analyses. Effective scientific assistants must therefore go beyond document retrieval and support evidence-grounded question answering, which motivates RAG systems that integrate high-quality retrieval, multi-step reasoning, and traceable answer generation.

In this work, we develop and evaluate an adaptive RAG agent over a muon collider scientific corpus. We employ a hybrid retriever combining dense semantic retrieval~\cite{karpukhin2020dpr,reimers2019sentencebert}, sparse BM25, and FAISS indexing~\cite{johnson2019faiss}, fused via RRF. On top of this retrieval backbone, a lightweight agent performs query decomposition and follow-up query generation to recover evidence missed by initial retrieval, while maintaining the fidelity and traceability required for scientific question answering in muon collider research.

The key contributions of this work are summarized as follows:

\begin{itemize}

\item We construct a comprehensive muon collider literature corpus and RAG benchmark, comprising a retrieval benchmark with expert-curated relevance annotations and an answer-level benchmark with reference answers, required key points, and unanswerable 
questions.

\item We propose an agentic hybrid RAG framework combining hybrid retrieval, agentic query decomposition, evidence expansion, and evidence-grounded answer generation, designed to maintain traceability between generated claims and supporting literature 
evidence.

\item We conduct comprehensive evaluations across retrieval and answer generation tasks, demonstrating that agentic hybrid RAG consistently outperforms representative baselines in retrieval effectiveness, answer quality, evidence coverage, and factual grounding. We find that hybrid retrieval and controlled evidence expansion are the primary drivers 
of these gains, validating the evidence-aware design of the framework.

\item We establish an end-to-end workflow spanning corpus construction, benchmark development, retrieval, reasoning, and answer generation, illustrating how evidence-aware RAG systems can support future muon collider studies and HEP analysis 
agents.

\end{itemize}
\section{Background and Motivation}
\subsection{RAG for HEP}
RAG integrates user queries with an external knowledge base by coupling two core components: a retrieval module and a generation module. The retrieval module selects relevant documents from the external corpus based on the input query, aiming to surface evidence most pertinent to the user’s information need. The generation module then conditions on the retrieved content to produce coherent and contextually grounded responses, leveraging a language model to synthesize information from the provided evidence.

In canonical RAG formulations, retrieval typically returns a fixed number of documents ranked by relevance to the query, and the generation process is constrained to rely primarily on this retrieved context. A widely used instantiation of this paradigm is dense retrieval with text embeddings, where documents are retrieved based on proximity to the query in a vector space, with similarity serving as a proxy for semantic relevance~\cite{gao2024retrievalaugmentedgenerationlargelanguage}. Although alternative retrieval strategies exist, this embedding-based formulation is often referred to as naive or standard vector-based RAG, and serves as a common baseline in many retrieval-augmented systems.

RAG-based question answering systems are increasingly being developed for applications in nuclear and high-energy physics. Recent work has explored retrieval and question answering over Electron–Ion Collider (EIC) literature~\cite{suresh2024eicrag,jat2026eicqa}, as well as within large LHC experiments such as LHCb~\cite{McGreivy:2025rrz} and CMS~\cite{Mallampalli:2026hrl}. These efforts are supported by a broader ecosystem of scientific text-mining and information infrastructure, including INSPIRE, S2ORC, SciBERT, and GROBID~\cite{jain2020s2orc,beltagy2019scibert,lopez2009grobid}. Collectively, they demonstrate the feasibility of applying retrieval-augmented methods to domain-specific scientific corpora, where relevant knowledge is distributed across large, heterogeneous, and rapidly evolving collections of literature.

Meanwhile, these studies also highlight limitations in current evaluation practices for scientific question answering. While general-purpose metrics and frameworks such as RAGAS, ROUGE, and classical information-retrieval measures~\cite{es2023ragas,lin2004rouge,voorhees1999trec,jarvelin2002ndcg} provide useful proxies for retrieval and generation quality, they do not fully capture the requirements of scientific settings. In particular, scientific question answering requires fine-grained assessment of evidence grounding, chunk-level retrieval correctness, and the ability to abstain when supporting evidence is absent, all of which are only partially reflected by standard automatic metrics. To address these requirements, we construct a dedicated evaluation framework that combines classical retrieval metrics with LLM-as-a-judge assessment for answer quality. The benchmark design and evaluation are described in Section~\ref{sec:experiments}, with detailed metric definitions provided in Appendix~\ref{appendix:retrieval_metrics} and~\ref{appendix:answer_metrics}.

\subsection{Promoting Muon Collider Analysis with RAG}

Muon colliders have re-emerged as a compelling option for future energy-frontier particle physics because they combine a lepton-collider initial state with access to multi-TeV center-of-mass energies. Compared with electrons, muons are heavier by a factor of about 200, which strongly suppresses synchrotron radiation and makes circular lepton-collider concepts feasible at energies where electron rings become impractical. As a result, the physics program is qualitatively distinct from both hadron and lower-energy lepton colliders, offering a clean initial state, high partonic center-of-mass energy, and sensitivity to both direct production of new particles and precision deviations from the Standard Model~\cite{muoncollider_pbc2020,black2024forum,imcc2022,accettura2024imcc, Qian:2021ihf, Jiang:2023mte,Jiang:2024wwa}. The physics case spans Higgs, electroweak, top, flavour, and beyond-the-Standard-Model studies, with particular emphasis on Higgs precision measurements and high-energy vector-boson fusion and multi-boson processes~\cite{barger1995higgs,Neuffer:2013wrd,celada2024higgs,costantini2020vectorboson,han2020higgs,abbott2023anomalous}.

The same features that enable this physics reach also introduce significant accelerator and detector challenges. Muon decays along the beam line produce intense beam-induced backgrounds, generating large fluxes of secondary particles with broad spatial and temporal distributions. These backgrounds increase detector occupancies, contaminate calorimeter energy deposits, and degrade tracking and reconstruction performance. Consequently, the machine-detector interface becomes a central design element rather than a peripheral engineering consideration. Mitigation strategies include dedicated shielding, optimized interaction-region design, precise timing, high-granularity detectors, and reconstruction algorithms capable of rejecting out-of-time activity~\cite{mokhov2011mdi,bartosik2020performance,collamati2021bib,lucchesi2020detector,accettura2024imcc}.

Beyond its physics motivation and detector challenges, the muon collider domain provides a useful testbed for retrieval-augmented generation systems in scientific research. Unlike established high-energy physics subfields, it lacks mature AI-assisted literature navigation tools and standardized benchmarks for information retrieval and synthesis. At the same time, relevant knowledge is distributed across accelerator design, beam-induced background studies, detector performance, and physics analyses, making it difficult to answer focused questions without cross-document evidence aggregation. This combination makes it well-suited for evaluating whether RAG systems can perform fine-grained retrieval, multi-source synthesis, and evidence-grounded reasoning under realistic scientific conditions.

To characterize the structure of information needs in this domain, we define representative application scenarios for our agentic hybrid RAG framework in Table~\ref{tab:applications}. These scenarios span accelerator design and beam performance, machine--detector interface studies, beam-induced background mitigation, Higgs and electroweak measurements, beyond-the-Standard-Model searches, multi-boson processes, and answerability checking.

\begin{table}[htbp]
\centering
\caption{Representative RAG application scenarios and query intents for muon collider analysis.}
\label{tab:applications}
\renewcommand{\arraystretch}{1.3}
\begin{tabular}{p{0.22\linewidth}p{0.35\linewidth}p{0.35\linewidth}}
\toprule
Query category & User intent & Retrieval objective \\
\midrule

Physics motivation &
Understand the motivation for muon colliders in future collider programs. &
Retrieve high-level collider comparisons and physics case discussions. \\

Beam cooling and beam quality &
Understand cooling requirements, emittance evolution, and beam-quality limitations. &
Retrieve accelerator concepts, cooling schemes, and beam-performance studies. \\

Machine--detector interface &
Relate luminosity goals, beam constraints, and interaction-region design. &
Retrieve accelerator design principles and machine--detector interface studies. \\

Beam-induced backgrounds &
Analyze the origin of beam-induced backgrounds and evaluate mitigation strategies. &
Retrieve detector-performance studies and simulation-based mitigation results. \\

Higgs and electroweak physics &
Summarize Higgs production channels and precision electroweak measurements. &
Integrate phenomenology with detector and collider constraints. \\

Beyond-Standard-Model physics &
Assess discovery potential and signatures of new physics scenarios. &
Retrieve BSM phenomenology, sensitivity studies, and collider projections. \\

Multi-boson processes &
Distinguish diboson, triboson, and EFT-driven anomalous interactions. &
Retrieve high-energy scattering and EFT interpretation literature. \\

Answerability check &
Verify whether claims are supported by available evidence. &
Prefer abstention over unsupported or hallucinated answers. \\

\bottomrule
\end{tabular}
\end{table}

These characteristics create a practical challenge for literature navigation. Scientific questions often require connecting accelerator constraints, beam-induced background mitigation, detector performance limitations, and physics analysis requirements across multiple sources. Correct answers frequently depend on retrieving specific evidence fragments rather than entire documents. For instance, a query on timing cuts for beam-induced background rejection requires different evidence from one on vector-boson fusion sensitivity to anomalous quartic gauge couplings, even though both arise within the same literature corpus. This motivates a retrieval-augmented system capable of chunk-level evidence retrieval, source traceability, and grounded answer synthesis for detector and physics studies.
\section{Methodology}\label{sec:method}

\subsection{Hybrid Retrieval}\label{sec:method_hybrid}

The retrieval stage combines sparse lexical retrieval and dense semantic retrieval. Sparse retrieval preserves exact technical terminology commonly used in HEP literature, while dense retrieval improves robustness to paraphrased scientific descriptions.

\paragraph{Sparse retriever.}
The sparse component uses BM25~\cite{robertson-2009}, which scores a tokenized query $q$ and document chunk $d$ according to term frequency, inverse document frequency, and length normalization:
\begin{equation}
S_{\rm BM25}(q,d)=
\sum_{t\in q}
\mathrm{IDF}(t)
\frac{f(t,d)(k_1+1)}
{f(t,d)+k_1(1-b+b|d|/\overline{|d|})}.
\end{equation}
BM25 prioritizes exact term overlap, making it particularly effective in HEP literature where key concepts are often expressed through stable acronyms and technical keywords such as BIB, MDI, VBS, and aQGC. The parameter $b \in [0,1]$ controls document-length normalization, interpolating between no normalization ($b=0$) and full normalization ($b=1$); we use $b=0.75$ as a standard compromise that mitigates bias toward longer chunks while remaining sensitive to length variation. Despite its precision on keyword-driven queries, BM25 is fundamentally limited by its reliance on exact lexical overlap: it cannot capture semantic similarity when relevant concepts are expressed using different terminology—such as "beam-induced background" versus "backgrounds from muon decays"—nor model compositional meaning or contextual relationships, motivating the complementary use of dense retrieval.

\paragraph{Dense retriever.}
The dense component embeds both queries and document chunks into a shared vector space using \texttt{sentence-transformers/all-MiniLM-L6-v2}. Chunk embeddings are indexed with FAISS under cosine similarity~\cite{gao2024retrievalaugmentedgenerationlargelanguage}.

Formally, let $f(\cdot)$ denote the encoder that maps a text input into an embedding vector:
\begin{equation}
\mathbf{e}_q = f(q), \quad \mathbf{e}_d = f(d).
\end{equation}

Dense retrieval ranks document chunks by cosine similarity:
\begin{equation}
S_{\text{dense}}(q,d) =
\frac{\mathbf{e}_q \cdot \mathbf{e}_d}
{\|\mathbf{e}_q\| \, \|\mathbf{e}_d\|}.
\end{equation}

Equivalently, FAISS performs nearest neighbor search in the embedding space:
\begin{equation}
d^* = \arg\max_{d \in \mathcal{D}} S_{\text{dense}}(q,d).
\end{equation}

Dense retrieval captures conceptual similarity even when surface forms differ significantly—for example, "beam-induced background" may be retrieved via "backgrounds from muon decays," and "power efficiency" queries may match chunks framed as "luminosity-normalized energy consumption." It also generalizes across heterogeneous writing styles across papers, detector studies, and phenomenology analyses, and improves recall for exploratory queries where users may not know the exact terminology in the literature. However, embedding-based similarity can over-generalize on fine-grained technical queries, and dense models may underperform on rare acronyms or newly introduced terminology not well represented in the embedding space. These limitations motivate combining dense retrieval with BM25 in a hybrid scheme.

\paragraph{Hybrid retriever.}
The sparse and dense rankings are merged using the weighted
reciprocal-rank fusion (RRF) score:
\begin{equation}
S_{\rm RRF}(c)=
\frac{w_d}{K+r_d(c)}
+
\frac{w_s}{K+r_s(c)},
\label{eq:rrf}
\end{equation}
where $r_d(c)$ and $r_s(c)$ denote the dense and sparse ranks of
chunk $c$, and $w_d$, $w_s$ are the corresponding weights.
We set $K=60$ following the original RRF formulation of Cormack et al.
~\cite{cormack2009reciprocal}, in which this value was empirically
shown to yield robust performance across diverse retrieval
benchmarks and has since been widely adopted as the de facto
standard in hybrid retrieval systems~\citep{pradeep2021rrf102,
askeda2024}.
The constant $K$ acts as a smoothing term that mitigates the
outsized influence of the single top-ranked candidate and prevents
any one retriever from dominating the fused score when it produces
a high-confidence outlier at rank~$1$.
We further constrain $w_d + w_s = 1$, so the fusion reflects a
convex combination of dense and sparse signals, improving
interpretability and reducing the effective hyperparameter space
to a single degree of freedom.
The default configuration uses $w_d=0.9$ and $w_s=0.1$,
reflecting the stronger semantic coverage of dense retrieval,
with the sparse retriever serving primarily as a high-precision
fallback for exact terminology, acronyms, and named entities.

\subsection{Agentic Query Decomposition}\label{sec:query_decomp}

For complex scientific queries that a single retrieval call cannot adequately address, we introduce an agentic query expansion layer built on top of the hybrid retriever. The core idea is to decompose the original query into a set of targeted subqueries, each probing a complementary aspect of the underlying information need, and to aggregate the resulting evidence into an expanded candidate pool.

\paragraph{Query decomposition.}
Given an input query $q$, the system applies three lightweight language-model prompts in sequence to produce a set of retrieval-oriented subqueries $\{q_1, q_2, \ldots, q_N\}$.

First, a domain tagging prompt identifies which physics domains are semantically relevant to $q$ from a controlled vocabulary (higgs, VBS, multiboson, detector, machine, BSM, ...), relying on semantic inference rather than keyword matching to handle paraphrase and implicit context.

Second, a query classification prompt assigns $q$ to one of three retrieval strategies: precise fact (a specific number, parameter, or direct claim), broad synthesis (a summary spanning multiple papers or concepts), or reasoning (causal, comparative, or mechanistic questions).

Third, a subquery generation prompt produces the final expansion conditioned on the original query, its detected tags, and its query type. Up to five additional subqueries are generated according to the classification: precise fact queries receive at most two narrow expansions to preserve retrieval precision, whereas reasoning queries are decomposed along mechanism, motivation, and limitation dimensions, and broad synthesis queries are split by domain or process boundary. Each subquery is self-contained and independently retrievable, targeting a specific facet of $q$ rather than paraphrasing it holistically. The prompt explicitly prohibits inventing paper titles, numerical values, or unsupported claims, ensuring that subqueries remain grounded in the original question. 

In this work, we use GPT-OSS-120B for all three stages, and the maximum subquery budget is capped at $N_{\max}=5$. The full prompt templates are provided in Appendix~\ref{app:prompts}.

\paragraph{Subquery retrieval and aggregation.}
Each generated subquery $q_i$ is passed through the same hybrid retrieval pipeline described in Sec.~\ref{sec:method_hybrid}, producing a ranked list of candidate chunks. The resulting $N$ candidate lists are merged by taking the union of retrieved chunks and deduplicating by chunk identifier. The final evidence pool retains the top-$M$ chunks ranked by their best RRF score across all subquery retrievals, where $M$ is the total evidence budget shared with the answer generator. This design ensures that the agentic expansion layer reuses the same sparse-dense retrieval infrastructure without introducing additional retrieval mechanisms and that evidence quality is governed by the same hybrid ranking criteria as the retrieval stage.

\subsection{Evidence-Grounded Answer Generation}

The answer generation module receives the original query together with a consolidated evidence set retrieved via both the original query and its decomposed subqueries. The generator is instructed to produce responses strictly grounded in the provided chunks, to cite supporting evidence, and to abstain when the retrieved material is insufficient to support a reliable answer. This grounding constraint is particularly important in detector and physics applications, where unsupported claims regarding background levels, detector performance, or physics reach can lead to incorrect scientific interpretations.

For each subquery $q_i$, relevant chunks are retrieved using the hybrid retrieval system described in Sec.~\ref{sec:method_hybrid}. The resulting candidate sets are then merged into a unified evidence pool through deduplication at the chunk level. When multiple subqueries retrieve overlapping or redundant evidence, duplicates are removed while preserving the highest-scoring occurrence according to the hybrid ranking function. The final evidence set is constrained to a fixed budget of top-$M$ chunks, ensuring a consistent input size for downstream generation. 

The language model then conditions on this evidence set to produce the final answer to the original query.  For benchmark evaluation, the same generation protocol is applied uniformly to both answerable and unanswerable queries, enabling consistent measurement of groundedness and abstention behavior, as detailed in Sec.~\ref{sec:experiments}.

\subsection{System Overview}
\begin{figure}[htbp]
\centering
\includegraphics[width=0.95\linewidth]{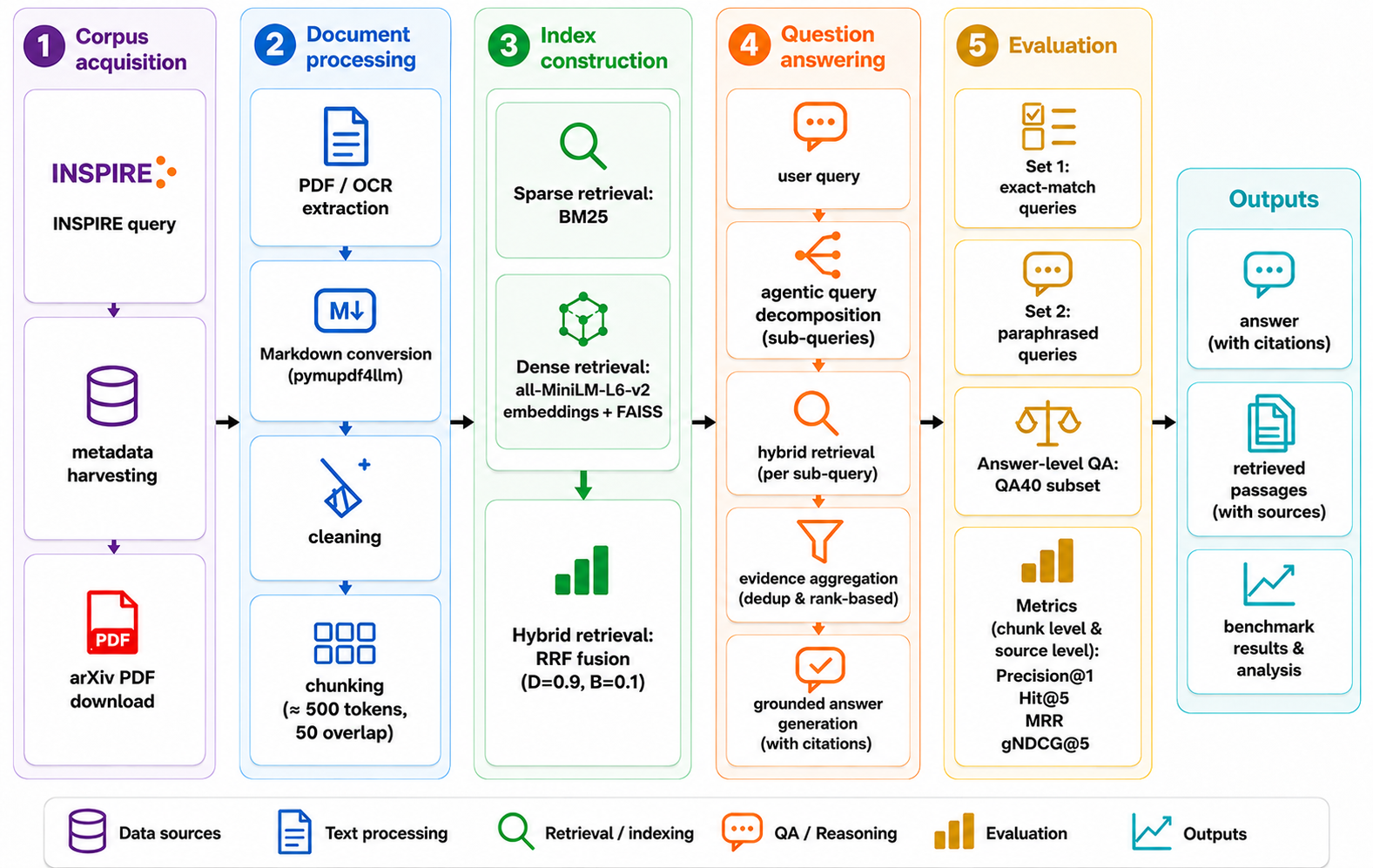}
\caption{Overview of the proposed agentic hybrid RAG system for scientific queries.}\label{fig:framework}
\end{figure}

Figure~\ref{fig:framework} presents an overview of the proposed system. The pipeline is organized into three tightly coupled stages: agentic query decomposition, hybrid retrieval, and evidence aggregation, followed by grounded answer generation.

Given a scientific query, the system first applies an agentic query decomposition module that generates a set of decomposed subqueries. The purpose of this stage is to explicitly expand the original information need into multiple complementary retrieval perspectives, particularly for complex scientific questions that involve multiple physical processes, detector effects, or analysis assumptions. Instead of relying on a single query formulation, this decomposition step increases coverage over heterogeneous evidence sources and reduces the risk of missing relevant chunks due to
lexical or semantic mismatch.

In the second stage, each decomposed subquery is independently processed by the hybrid retrieval module. This module combines BM25-based sparse retrieval, which is effective for exact matching of technical terminology and acronyms, with dense embedding-based retrieval, which captures semantic similarity across paraphrased or contextually reformulated scientific descriptions. The same hybrid retrieval pipeline is also applied to the original query, ensuring that the system retains a high-precision evidence backbone while simultaneously exploring expanded query views. Prior to retrieval, the scientific document corpus is segmented into chunk-level chunks, which are indexed under both the sparse and dense representations. The result of this stage is a set of ranked chunk lists corresponding to the original query and each decomposed subquery.

In the third stage, all retrieved chunks are merged into a unified evidence pool. This aggregation process performs deduplication at the chunk level to remove redundant content retrieved from multiple query views, while preserving the highest-ranked occurrence of each chunk according to the hybrid scoring signal. A fixed evidence budget is then enforced through rank-based selection, which balances relevance and diversity while maintaining a compact and controllable context size for downstream generation.

Finally, the answer generation module conditions on the consolidated evidence set to produce the final response. The model is explicitly constrained to ground its output in the retrieved chunks, refusing to synthesize claims that lack direct support in the evidence pool. This end-to-end design integrates decomposition-driven query expansion, hybrid sparse-dense retrieval, and evidence fusion into a unified framework that improves both recall and robustness in scientific question answering.
\section{Experiments}\label{sec:experiments}

\subsection{Experimental Setup and Benchmarks}
\paragraph{Corpus and benchmark.}
The collected corpus contains 219 muon collider publications and is segmented into 5,813 indexed chunks, which serve as the fundamental retrieval units throughout this work. To support evidence attribution and source traceability, each chunk retains its associated metadata. For dense retrieval, all chunks are encoded into 384-dimensional embeddings and indexed using FAISS. 

The benchmark consists of two components: a retrieval benchmark and an answer-generation benchmark, designed to decouple retrieval quality from end-to-end generation performance. The following subsections introduce each component in detail.

\paragraph{Retrieval benchmark and metrics.}
The retrieval benchmark consists of 58 questions, including 45 retrievable and 13 unretrievable questions, summarized in Table~\ref{tab:retriever_benchmark}. 
Retrieval performance is evaluated on the retrievable subset, while the unretrievable questions are reserved for abstention evaluation and robustness analysis.

\begin{table}[H]
\centering
\caption{Retrieval benchmark overview.}
\label{tab:retriever_benchmark}
\begin{tabular}{lcc}
\toprule
Component & Count & Description \\
\midrule
Total benchmark questions & 58 & Complete evaluation set \\
Retrievable questions & 45 & Used for retrieval evaluation \\
Unretrievable questions & 13 & Used for abstention evaluation \\
\bottomrule
\end{tabular}
\end{table}

For each retrievable question, candidate chunks are annotated with graded relevance judgments on a four-level scale. Grade 3 denotes chunks that directly answer the question, grade 2 indicates strong supporting evidence, grade 1 corresponds to relevant contextual information, and grade 0 denotes non-relevant content.

For retrieval evaluation, we report chunk-level Precision@1, Hit@5, Mean Reciprocal Rank (MRR), and graded Normalized Discounted Cumulative Gain (gNDCG)@5. The first three metrics treat relevance as binary, whereas gNDCG@5 leverages the graded relevance judgments defined above to evaluate ranking quality. The formulations of retrieval metrics are detailed in Appendix~\ref{appendix:retrieval_metrics}.

In addition, we report source-level retrieval performance, which evaluates whether at least one chunk originating from a relevant source document appears within the retrieved top-k results. 
This complementary metric reflects the practical utility of retrieval for literature discovery, where locating the correct paper may be sufficient even when the exact evidence chunk is not ranked highest.

\paragraph{End-to-end answer generation benchmark and metrics.}
The answer generation benchmark is intentionally distinct from the retrieval benchmark, enabling assessment of the complete retrieval-to-generation pipeline rather than the retrieval performance in isolation.
Retrieval quality alone does not adequately reflect end-to-end answer generation performance.
We construct a dedicated answer-level benchmark consisting of 40 questions, including 35 answerable and 5 unanswerable cases, summarized in Table~\ref{tab:answer_benchmark}.
Each question is annotated with a reference answer, required key points, and a list of unsupported claims that should not appear in the response.
The benchmark evaluates answer correctness, evidence coverage, hallucination resistance, and abstention behavior, complementing retrieval-focused evaluation.

\begin{table}[H]
\centering
\caption{Answer generation benchmark overview.}
\label{tab:answer_benchmark}
\begin{tabular}{lcc}
\toprule
Component & Count & Description \\
\midrule
Total benchmark questions & 40 & Complete QA evaluation set \\
Answerable questions & 35 & Used for correctness and key-point coverage evaluation \\
Unanswerable questions & 5 & Used for abstention evaluation \\
Required key points & 197 & Reference-answer evaluation rubric \\
\bottomrule
\end{tabular}
\end{table}

To evaluate the generated answers, a deterministic LLM-as-a-judge prompt compares each generated answer against the reference answer, required key points, and unsupported-claim criteria.
We report Good Rate, Satisfactory-or-Better Rate, Key-Point Coverage, Hallucination Rate, and Abstention Accuracy. We also perform qualitative inspection of representative examples to validate the consistency of automated judgments. The formulations of answer-generation metrics are detailed in Appendix~\ref{appendix:answer_metrics}.

\subsection{Hybrid Retriever Optimization}

\begin{figure}[htbp]
\centering
\includegraphics[width=0.86\linewidth]{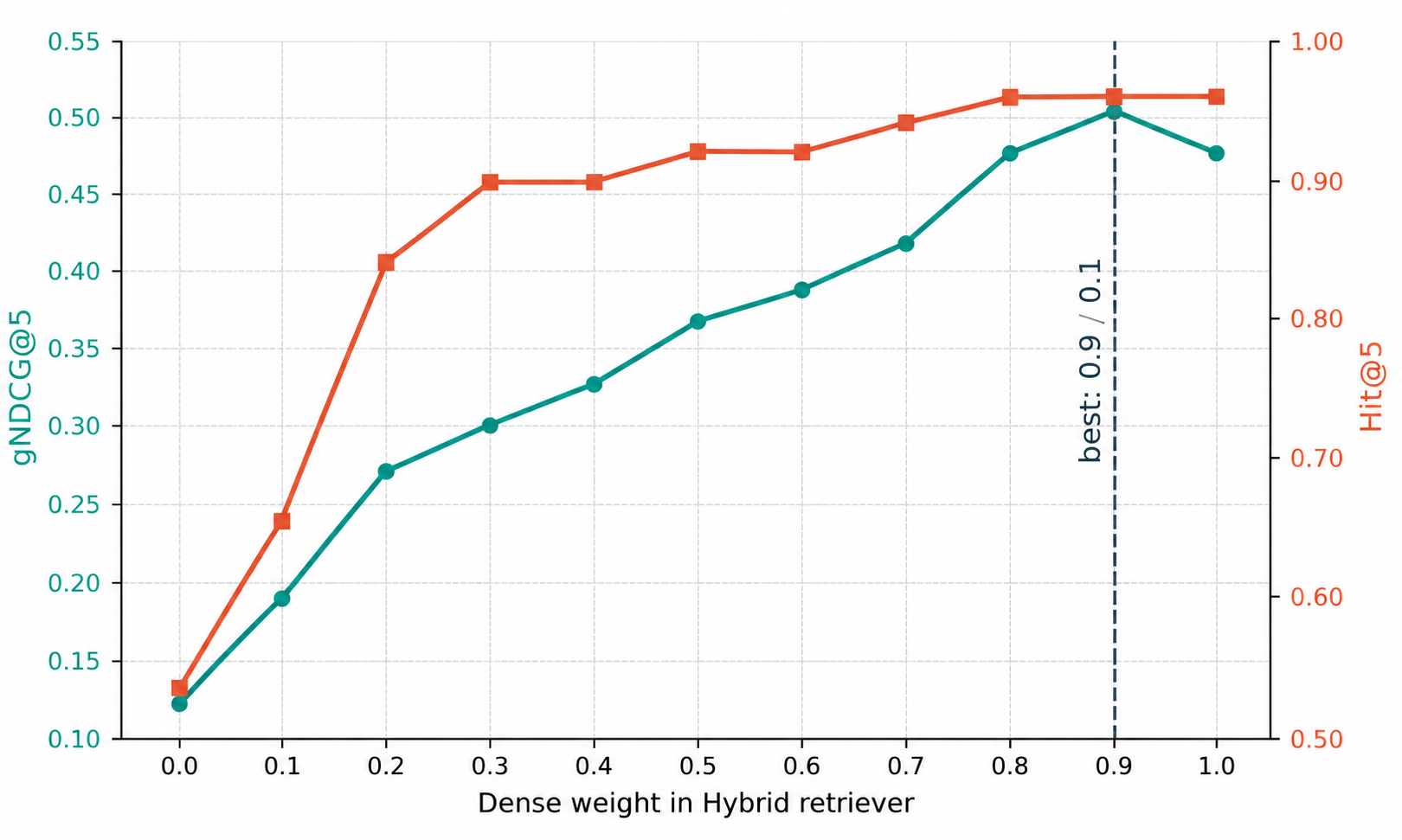}
\caption{Dense weight optimization of the hybrid retriever.}
\label{fig:sweep}
\end{figure}

Figure~\ref{fig:sweep} presents the weight optimization study for the hybrid retriever, where dense semantic retrieval and sparse lexical retrieval are combined with different relative weights.

The results indicate that retrieval performance is strongly influenced by the dense retriever component. Increasing the weight of semantic retrieval generally improves ranking quality, while a small lexical component remains beneficial. The best performance is achieved with a dense-retrieval weight of 0.9 and a sparse-retrieval (BM25) weight of 0.1, which is therefore adopted as the default configuration in all subsequent experiments.

This behavior is consistent with the characteristics of the benchmark. Many questions require semantic matching across heterogeneous descriptions of detector concepts, physics processes, and analysis methodologies, which favors dense retrieval. At the same time, a modest BM25 component helps preserve sensitivity to specialized high-energy physics terminology, detector component names, and exact technical expressions. Overall, the hybrid retriever combines the broad semantic coverage of dense retrieval with the lexical precision of BM25, outperforming either signal alone.

\subsection{Retrieval Evaluation}

\begin{table}[htbp]
\centering
\caption{Chunk-level retrieval performance on the retrievable set of questions. Higher values indicate better performance, and the best score in each column is highlighted in \textbf{bold}.}
\label{tab:retrieval}
\begin{tabular}{lcccc}
\toprule
Retriever & Precision@1 & Hit@5 & MRR & gNDCG@5 \\
\midrule
Sparse (BM25) & 0.222 & 0.467 & 0.351 & 0.122 \\
Dense (vector) & 0.689 & \textbf{0.956} & 0.806 & 0.484 \\
Hybrid (RRF) & \textbf{0.756} & \textbf{0.956} & \textbf{0.843} & \textbf{0.510} \\
\bottomrule
\end{tabular}
\end{table}

\begin{figure}[htbp]
\centering
\includegraphics[width=0.9\linewidth]{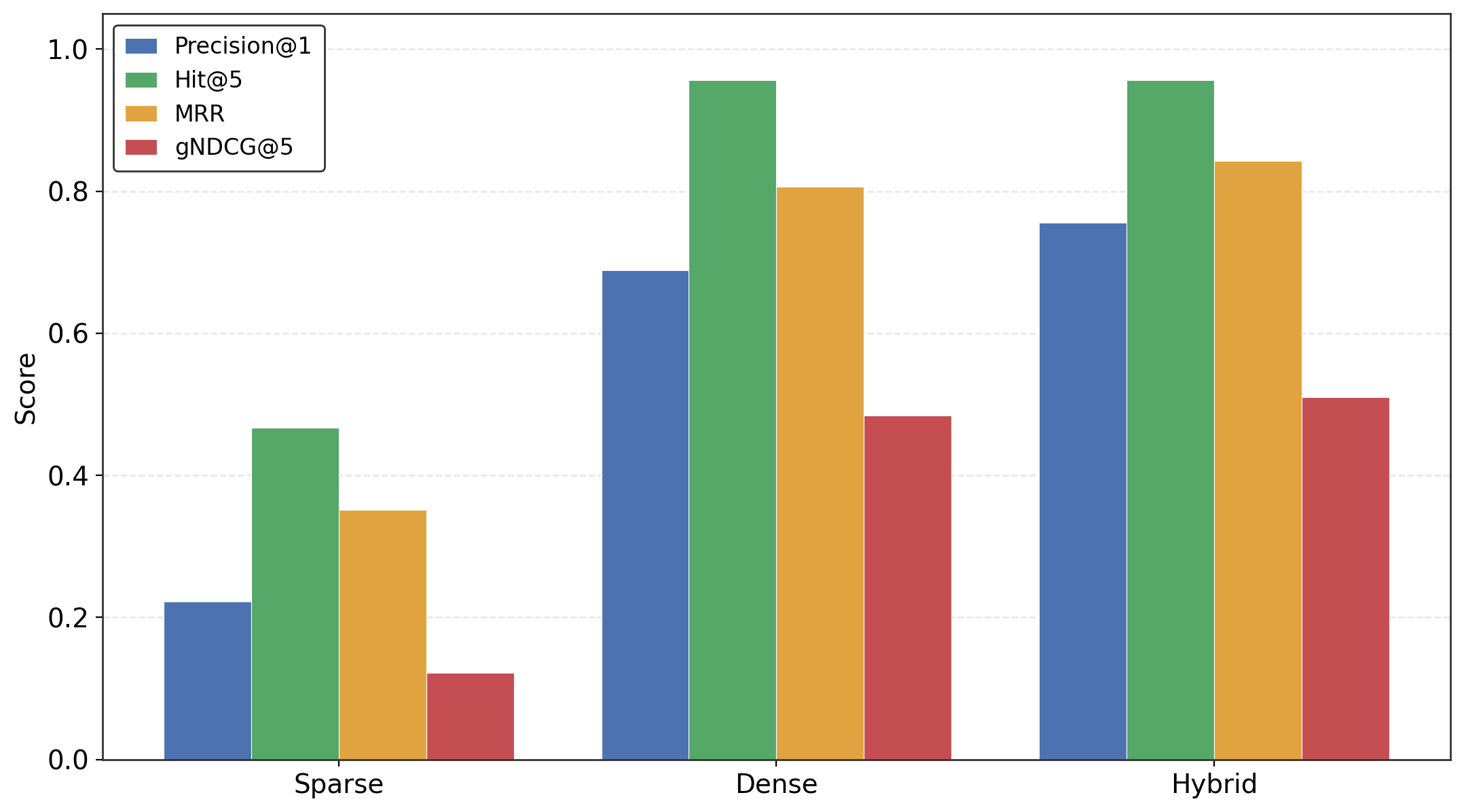}
\caption{Chunk-level retrieval performance across different retrievers on the retrievable questions.}
\label{fig:retrieval_evaluation}
\end{figure}

Because the agentic workflow performs query decomposition and retrieves evidence for multiple sub-queries, it is not directly comparable under fixed single-query retrieval metrics. Therefore, it is excluded from the chunk-level retrieval evaluation and is instead assessed in the end-to-end answer-generation setting.

Table~\ref{tab:retrieval} and Figure~\ref{fig:retrieval_evaluation} present the chunk-level retrieval performance of the three retrievers. The hybrid approach achieves the strongest overall performance, outperforming both standard BM25 and dense vector retrieval. BM25 alone underperforms due to the prevalence of semantically diverse queries that require conceptual matching beyond exact lexical overlap, whereas dense retrieval captures most semantic signals. By combining dense and sparse retrievals, the hybrid configuration provides both strong semantic matching and robustness to exact technical terminology.
\begin{figure}[htbp]
\centering
\includegraphics[width=0.9\linewidth]{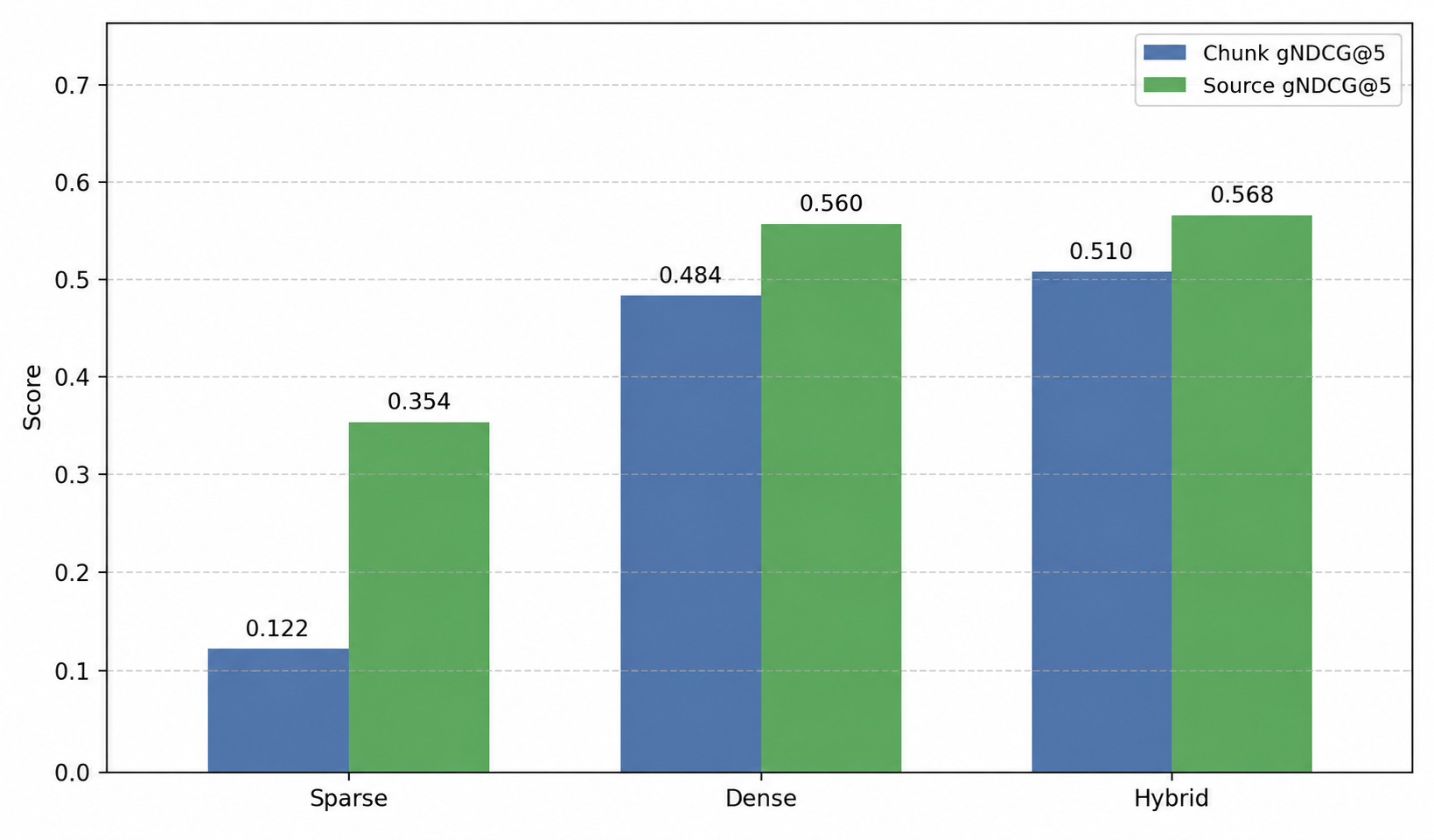}
\caption{Chunk-level vs. source-level retrieval performance.}
\label{fig:chunk_source}
\end{figure}

Figure~\ref{fig:chunk_source} compares chunk-level and source-level retrieval performance. Source-level performance is consistently higher than chunk-level performance, indicating that the system often retrieves the correct source document even when the exact annotated chunk is not ranked first. This reflects the fact that multiple chunks within the same document may provide redundant or complementary evidence for the same query, even if only one is explicitly annotated as relevant.

The retrieval evaluation establishes the hybrid retriever as the strongest retrieval backbone. We next evaluate whether agentic query decomposition and evidence aggregation improve answer quality when built on top of this retriever.

\subsection{Answer Generation Evaluation}

The retrieval evaluation establishes the hybrid retriever as the strongest retrieval backbone. We next evaluate whether agentic query decomposition and evidence aggregation improve answer quality when built on top of this retriever.

\begin{figure}[htbp]
\centering
\includegraphics[width=0.86\linewidth]{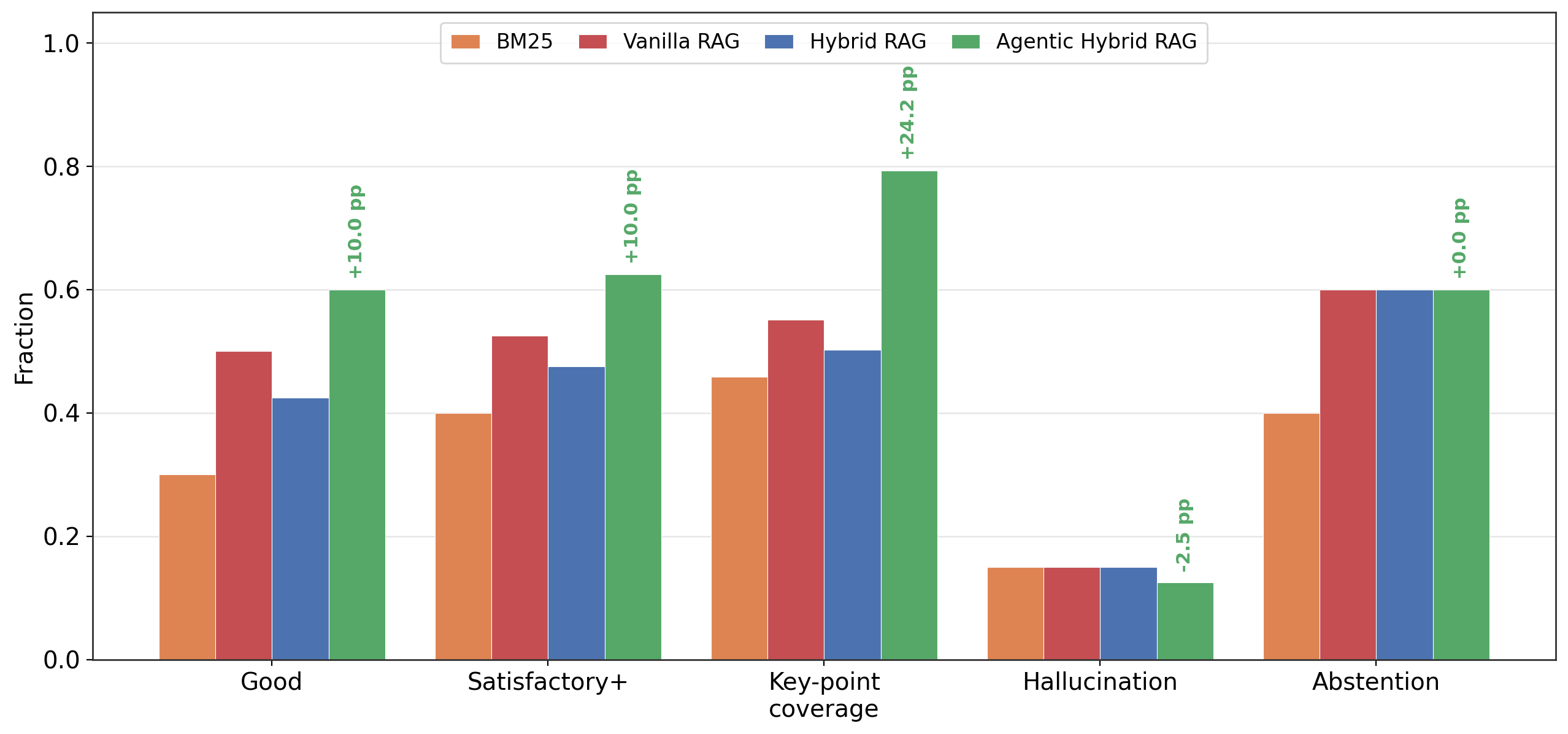}
\caption{Answer-generation performance across four methods on the 40-question benchmark. Numbers on Agentic Hybrid RAG show percentage-point improvement vs. Vanilla RAG.}
\label{fig:qa-improvement}
\end{figure}

Table~\ref{tab:qa-results} and Figure~\ref{fig:qa-improvement} summarize the answer-generation results on the 40-question benchmark. We compare BM25, vanilla RAG, hybrid RAG, and the proposed agentic hybrid RAG. GPT-OSS-120B is used for both answer generation and answer evaluation.

Evaluation is conducted using rubric-based metrics. Good and Satisfactory+ denote two levels of correctness based on reference answers and required key points. Key-point coverage measures the fraction of required reference points covered by the generated response. Hallucination rate denotes the proportion of answers containing unsupported claims, while the abstention rate measures the fraction of correctly unanswered questions. Higher values indicate better performance for all metrics except hallucination rate, where lower values are preferred.

\begin{table}[H]
\centering
\caption{Answer generation performance on the 40-question benchmark. Higher values are better for all metrics except hallucination rate (lower is better). The best score is highlighted in \textbf{bold}.}
\label{tab:qa-results}
\small
\setlength{\tabcolsep}{4pt}
\renewcommand{\arraystretch}{1.12}
\begin{tabularx}{\textwidth}{lccccc}
\toprule
Method & Good[\%] & Satisfactory+[\%] & Key-points[\%] & Hallucination[\%] & Abstention[\%] \\
\midrule
BM25 & 30.0 & 40.0 & 45.9 & 15.0 & 40.0\\
Vanilla RAG & 50.0 & 52.5 & 55.1 & 15.0 & \textbf{60.0} \\
Hybrid RAG & 42.5 & 47.5 & 50.2 & 15.0 & \textbf{60.0} \\
Agentic Hybrid RAG & \textbf{60.0} & \textbf{62.5} & \textbf{79.3} & \textbf{12.5} & \textbf{60.0} \\
\bottomrule
\end{tabularx}
\end{table}

Interestingly, improvements in retrieval metrics do not directly translate into answer-generation quality. Although the hybrid retriever achieves the strongest retrieval performance, the corresponding Hybrid RAG baseline does not outperform Vanilla RAG on this benchmark. This discrepancy suggests that retrieval quality alone is insufficient to characterize end-to-end answer quality, motivating the use of answer-level evaluation in addition to retrieval metrics.

Agentic hybrid RAG achieves the strongest overall answer-generation performance. Compared with Vanilla RAG, the Good rate increases from 50.0\% to 60.0\% and the Satisfactory-or-Better rate from 52.5\% to 62.5\%. The largest gain is observed in key-point coverage, which rises from 55.1\% to 79.3\%, indicating substantially more complete utilization of retrieved evidence. The hallucination rate is also reduced from 15.0\% to 12.5\%, suggesting that additional evidence aggregation does not increase unsupported claims.

Abstention accuracy remains unchanged at 60.0\% across retrieval-based methods, indicating that retrieval improvements alone are insufficient to fully address unsupported-question detection.

\section{Conclusion and Outlook}

This work introduced a benchmark for retrieval-augmented scientific question answering in the muon collider domain, covering detector and physics literature spanning accelerator concepts, beam-induced backgrounds, machine-detector interfaces, Higgs studies, multi-boson processes, vector-boson scattering, and detector-performance research. Building on this benchmark, we developed and evaluated an agentic hybrid RAG framework designed for evidence-grounded scientific literature exploration.

A key observation of this study is that agentic retrieval should complement, rather than replace, a strong hybrid retriever, particularly in answer generation settings. Across the retrieval benchmark, the hybrid retrieval component of agentic hybrid RAG consistently achieves the strongest performance among the evaluated retrieval methods. This indicates that agentic retrieval is most effective when built on a strong underlying retrieval backbone rather than used as a replacement.

This distinction is particularly important for HEP analysis agents, where retrieved evidence may directly influence detector studies, background estimation, and physics analysis workflows. In such settings, the value of an answer depends not only on its linguistic quality but also on the ability to trace every scientific claim back to supporting evidence.

The answer-level evaluation further demonstrates the value of controlled evidence expansion. Agentic hybrid RAG outperforms all evaluated baselines in answer quality, evidence coverage, and factual grounding, while maintaining strong citation fidelity and low hallucination rates. Overall, agentic reasoning is most effective when applied to evidence organization, contextualization, and answer synthesis, rather than as an unconstrained replacement for retrieval. More broadly, the study highlights the importance of balancing retrieval precision with reasoning flexibility in scientific RAG systems.

Several limitations remain. The benchmark is self-constructed, some degree of terminology overlap exists between queries and reference evidence, and answer evaluation relies partly on an LLM-as-a-judge framework. Future work should therefore incorporate community-reviewed benchmarks, broader domain coverage, and expert auditing of both retrieval and answer quality. Beyond benchmark evaluation, it will be important to assess the framework in realistic scientific workflows, including detector-background studies, machine-detector interface design reviews, detector optimization tasks, and physics-performance analyses. Such studies would provide a more direct measure of how evidence-grounded retrieval and reasoning can support day-to-day research activities.

Looking further ahead, practical deployment of HEP analysis agents will require more than strong retrieval and question answering. Future systems must operate over versioned scientific corpora, maintain persistent links between generated conclusions and supporting evidence, and provide transparent mechanisms for citation, verification, and human review. We view agentic hybrid RAG as a foundational component of this broader vision: an evidence-aware knowledge layer that enables future HEP analysis agents to retrieve, inspect, connect, and reason over scientific literature while keeping analysis decisions and scientific claims grounded in their sources.

\section*{Data availability}
The code for the agentic hybrid RAG framework is available at
\url{https://github.com/AItutorialjrb/RAG_muon_JINST}. The muon collider literature
corpus, the prebuilt retrieval indices, and the retrieval and answer-generation
benchmarks are available at
\url{https://huggingface.co/datasets/Ruobingjiang/AI_muon_collider_RAG}.

\appendix
\newpage
\appendix
\section*{Appendix}
\addcontentsline{toc}{section}{Appendix}

\section{Query Decomposition Prompt Templates}
\label{app:prompts}

The following prompts are used in the three-stage query decomposition
pipeline described in Section~\ref{sec:query_decomp}.
All prompts instruct the model to return structured JSON output only,
with no preamble or markdown formatting.

\subsection*{Stage 1: Domain Tag Detection}
\label{app:prompts:tag}

\begin{tcolorbox}[
  title=\texttt{TAG\_DETECTION\_PROMPT},
  fonttitle=\small\ttfamily,
  colback=gray!5, colframe=gray!40,
  breakable
]
\begin{verbatim}
You are a high-energy-physics query analysis assistant.
Given a user query about muon-collider physics, detector
studies, or related literature, identify the relevant domains.

Allowed tags:
  - higgs
  - multiboson
  - vbs
  - aqgc
  - detector
  - machine
  - general

Rules:
  - Return only tags that are semantically relevant.
  - Do not rely only on exact keywords; infer from context
    and paraphrase.
  - Use "general" only if no specific tag applies.
  - Return JSON only.

Output schema:
{
  "tags":   ["..."],
  "reason": "brief explanation"
}
\end{verbatim}
\end{tcolorbox}

\subsection*{Stage 2: Query Classification}
\label{app:prompts:classify}

\begin{tcolorbox}[
  title=\texttt{QUERY\_CLASSIFICATION\_PROMPT},
  fonttitle=\small\ttfamily,
  colback=gray!5, colframe=gray!40,
  breakable
]
\begin{verbatim}
You are classifying scientific RAG queries.
Classify the query into exactly one query type:

  - precise_fact:     asks for a specific number, result,
                      parameter, paper claim, definition,
                      or direct fact.
  - broad_synthesis:  asks for a summary across several
                      papers, topics, or concepts.
  - reasoning:        asks why, how, compare, connect,
                      affect, influence, limitation,
                      motivation, or implication.
  - paraphrase:       asks to rewrite, polish, or rephrase
                      provided text.

Rules:
  - Use semantic intent, not only keywords.
  - If the query asks for both a fact and an explanation,
    choose reasoning.
  - Return JSON only.

Output schema:
{
  "query_type": "...",
  "reason":     "brief explanation"
}
\end{verbatim}
\end{tcolorbox}

\subsection*{Stage 3: Subquery Generation}
\label{app:prompts:decomp}

\begin{tcolorbox}[
  title=\texttt{QUERY\_DECOMPOSITION\_PROMPT},
  fonttitle=\small\ttfamily,
  colback=gray!5, colframe=gray!40,
  breakable
]
\begin{verbatim}
You are a domain-aware query decomposition module for a
scientific RAG system.

Given:
  1. the original user query,
  2. detected domain tags,
  3. query type,
generate retrieval-oriented subqueries.

Rules:
  - The first subquery must be the original query verbatim.
  - Generate 2 to 6 additional subqueries only if they
    genuinely aid retrieval.
  - Subqueries should retrieve supporting evidence, not
    answer the question directly.
  - For precise_fact queries: keep subqueries narrow;
    generate at most 2 extra subqueries.
  - For reasoning queries: decompose into mechanism,
    motivation, limitation, and relevant evidence angles.
  - For broad_synthesis queries: decompose by domain
    or process boundary.
  - Do not invent paper titles, numerical values, or
    unsupported claims.
  - Avoid generic subqueries that would retrieve too many
    unrelated chunks.
  - Return JSON only.

Output schema:
{
  "subqueries": [
    "original query",
    "subquery 1",
    "subquery 2"
  ],
  "reason": "brief explanation"
}
\end{verbatim}
\end{tcolorbox}

\subsection*{Illustrative Example}
\label{app:prompts:example}

Table~\ref{tab:decomp_example} shows the pipeline output for a
representative \texttt{reasoning} query tagged as
\texttt{vbs} and \texttt{aqgc}.

\begin{table}[h]
\centering
\small
\begin{tabular}{p{0.1\linewidth} p{0.84\linewidth}}
\toprule
\textbf{Stage} & \textbf{Output} \\
\midrule
Input
  & \textit{``Why is a muon collider particularly sensitive
    to anomalous quartic gauge couplings in VBS?''} \\
\midrule
Tags
  & \texttt{vbs}, \texttt{aqgc} \\
\midrule
Type
  & \texttt{reasoning} \\
\midrule
\multirow{5}{*}{Subqueries}
  & 1.~Why is a muon collider particularly sensitive to
       anomalous quartic gauge couplings in VBS? \\
  & 2.~Vector boson scattering at a muon collider \\
  & 3.~Vector boson fusion high-energy muon collider \\
  & 4.~VBS anomalous quartic gauge couplings muon collider \\
  & 5.~Sensitivity to aQGC in VBS at high-energy lepton
       colliders \\
  & 6.~Unitarity and aQGC constraints from VBS \\
\bottomrule
\end{tabular}
\caption{Pipeline output for a \texttt{reasoning} query
  tagged \texttt{vbs} + \texttt{aqgc}.
  Subquery~1 is the original query verbatim; subqueries 2--6
  target mechanism, phenomenology, and theoretical context
  angles respectively.}
\label{tab:decomp_example}
\end{table}

\section{Retrieval Metrics Formulation}\label{appendix:retrieval_metrics}

Let $\mathcal{R}_k=\{d_1,\ldots,d_k\}$ denote the top-$k$ retrieved chunks for a query, ranked by decreasing retrieval score. Let $\mathrm{rel}_i$ denote the graded relevance score of the chunk at rank $i$.

\paragraph{Precision@k.}
Precision@k measures the fraction of retrieved chunks within the top-$k$ results that are relevant:
\begin{equation}
\mathrm{Precision}@k
=
\frac{1}{k}
\sum_{i=1}^{k}
\mathbf{1}(\mathrm{rel}_i>0),
\end{equation}
where $\mathbf{1}(\cdot)$ is the indicator function.

\paragraph{Hit@k.}
Hit@k evaluates whether at least one relevant chunk appears among the top-$k$ retrieved results:
\begin{equation}
\mathrm{Hit}@k
=
\mathbf{1}
\left(
\sum_{i=1}^{k}
\mathbf{1}(\mathrm{rel}_i>0)
>0
\right).
\end{equation}
For a set of queries, Hit@k is averaged across all queries.

\paragraph{Mean Reciprocal Rank (MRR).}
Let $r$ denote the rank position of the first relevant chunk. The reciprocal rank (RR) is defined as
\begin{equation}
\mathrm{RR}
=
\frac{1}{r}.
\end{equation}
The Mean Reciprocal Rank over a set of $N$ queries is
\begin{equation}
\mathrm{MRR}
=
\frac{1}{N}
\sum_{q=1}^{N}
\frac{1}{r_q},
\end{equation}
where $r_q$ is the rank of the first relevant chunk for query $q$.

\paragraph{Graded Discounted Cumulative Gain (gDCG).}
To account for graded relevance, the Discounted Cumulative Gain at rank $k$ is defined as
\begin{equation}
\mathrm{gDCG}@k
=
\sum_{i=1}^{k}
\frac{2^{\mathrm{rel}_i}-1}
{\log_2(i+1)}.
\end{equation}

\paragraph{Graded Normalized Discounted Cumulative Gain (gNDCG).}
The graded Normalized Discounted Cumulative Gain is obtained by normalizing gDCG with the ideal ranking:
\begin{equation}
\mathrm{gNDCG}@k
=
\frac{\mathrm{gDCG}@k}
{\mathrm{IDCG}@k},
\end{equation}
where $\mathrm{IDCG}@k$ denotes the maximum achievable gDCG obtained by sorting retrieved chunks according to decreasing relevance scores.

\section{Answer Metrics Formulation}\label{appendix:answer_metrics}

For answer generation evaluation, a deterministic judge prompt compares each generated answer against the reference answer, required key points, and unsupported-claim criteria. Let $N$ denote the total number of evaluated questions.

\paragraph{Good Rate.}
Good Rate measures the fraction of answers judged as \textit{Good}:
\begin{equation}
\mathrm{GoodRate}
=
\frac{1}{N}
\sum_{i=1}^{N}
\mathbf{1}(y_i=\mathrm{Good}),
\end{equation}
where $y_i$ is the judge label assigned to the $i$-th answer.

\paragraph{Satisfactory-or-Better Rate.}
This metric measures the fraction of answers judged as either \textit{Good} or \textit{Satisfactory}:
\begin{equation}
\mathrm{Sat+Rate}
=
\frac{1}{N}
\sum_{i=1}^{N}
\mathbf{1}
\left(
y_i \in
\{\mathrm{Good},\mathrm{Satisfactory}\}
\right).
\end{equation}

\paragraph{Key-Point Coverage.}
Let $K_i$ denote the set of required key points for question $i$, and let $\hat{K}_i$ denote the subset correctly covered by the generated answer. Key-Point Coverage is defined as
\begin{equation}
\mathrm{KPC}
=
\frac{1}{N}
\sum_{i=1}^{N}
\frac{|\hat{K}_i|}{|K_i|}.
\end{equation}

\paragraph{Hallucination Rate.}
Hallucination Rate measures the proportion of answers containing unsupported factual claims:
\begin{equation}
\mathrm{HallucinationRate}
=
\frac{1}{N}
\sum_{i=1}^{N}
\mathbf{1}(h_i=1),
\end{equation}
where $h_i=1$ indicates that the judge identifies at least one unsupported claim in the generated answer.

\paragraph{Abstention Accuracy.}
For questions labeled as unanswerable from the available evidence, Abstention Accuracy measures whether the model correctly refrains from providing unsupported answers:
\begin{equation}
\mathrm{AbstentionAcc}
=
\frac{1}{M}
\sum_{i=1}^{M}
\mathbf{1}(a_i=\hat{a}_i),
\end{equation}
where $M$ is the number of unanswerable questions, $a_i$ is the ground-truth abstention label, and $\hat{a}_i$ is the model's abstention decision.

In addition, qualitative inspection of representative examples is performed to verify the consistency of automated judgments and to identify common failure modes.

\section{Reproducibility Commands}

Primary retrieval benchmark:
\begin{lstlisting}[style=code]
python evaluate_agentic.py \
  --query_file data/eval/queries_eval_v4_template.json \
  --dense_weight 0.9 \
  --bm25_weight 0.1 \
  --sweep_weights \
  --output data/eval/eval_results_agentic_retrieval_FULL.json
\end{lstlisting}

Answer-level QA evaluation:
\begin{lstlisting}[style=code]
python evaluate_agentic.py \
  --query_file data/eval/queries_eval_qa40_publishable.json \
  --dense_weight 0.9 \
  --bm25_weight 0.1 \
  --judge_answers \
  --qa_systems hybrid,agentic,oracle \
  --resume \
  --output data/eval/eval_results_agentic_qa40_FULL.json
\end{lstlisting}

Figure and table generation:
\begin{lstlisting}[style=code]
python make_qa_paper_outputs.py \
  --input data/eval/eval_results_agentic_qa40_FULL.json \
  --outdir paper_outputs_qa40_FULL
\end{lstlisting}

\bibliographystyle{JHEP}
\bibliography{main}

@article{lewis2020rag,
  title={Retrieval-Augmented Generation for Knowledge-Intensive NLP Tasks},
  author={Lewis, Patrick and Perez, Ethan and Piktus, Aleksandra and Petroni, Fabio and Karpukhin, Vladimir and Goyal, Naman and K{\"u}ttler, Heinrich and Lewis, Mike and Yih, Wen-tau and Rockt{\"a}schel, Tim and others},
  journal={Advances in Neural Information Processing Systems},
  volume={33},
  pages={9459--9474},
  year={2020}
}

@inproceedings{reimers2019sentencebert,
  title={Sentence-BERT: Sentence Embeddings using Siamese BERT-Networks},
  author={Reimers, Nils and Gurevych, Iryna},
  booktitle={Proceedings of EMNLP-IJCNLP},
  pages={3982--3992},
  year={2019}
}

@article{johnson2019faiss,
  title={Billion-scale similarity search with GPUs},
  author={Johnson, Jeff and Douze, Matthijs and J{\'e}gou, Herv{\'e}},
  journal={IEEE Transactions on Big Data},
  volume={7},
  number={3},
  pages={535--547},
  year={2019}
}

@inproceedings{karpukhin2020dpr,
  title={Dense Passage Retrieval for Open-Domain Question Answering},
  author={Karpukhin, Vladimir and Oguz, Barlas and Min, Sewon and Lewis, Patrick and Wu, Ledell and Edunov, Sergey and Chen, Danqi and Yih, Wen-tau},
  booktitle={Proceedings of EMNLP},
  pages={6769--6781},
  year={2020}
}

@article{singh2025agenticrag,
  title={Agentic retrieval-augmented generation: A survey on agentic rag},
  author={Singh, Aditi and Ehtesham, Abul and Kumar, Saket and Khoei, Tala Talaei and Vasilakos, Athanasios V},
  journal={arXiv preprint arXiv:2501.09136},
  year={2025}
}

@article{suresh2024eicrag,
   author = "Suresh, Karthik and Kackar, Neeltje and Schleck, Luke and Fanelli, Cristiano",
    title = "{Towards a RAG-based summarization for the Electron Ion Collider}",
    eprint = "2403.15729",
    archivePrefix = "arXiv",
    primaryClass = "cs.CL",
    doi = "10.1088/1748-0221/19/07/C07006",
    journal = "JINST",
    volume = "19",
    number = "07",
    pages = "C07006",
    year = "2024"
}

@article{jat2026eicqa,
  title={Retrieval-Augmented Question Answering over Scientific Literature for the Electron-Ion Collider},
  author={Jat, Tina J. and Ghosh, T. and Suresh, Karthik},
  journal={arXiv preprint arXiv:2604.02259},
  year={2026}
}

@article{barger1995higgs,
  author = "Barger, Vernon D. and Berger, M. S. and Gunion, J. F. and Han, Tao",
    title = "{S-channel Higgs boson production at a muon muon collider}",
    eprint = "hep-ph/9504330",
    archivePrefix = "arXiv",
    reportNumber = "UCD-95-12, MAD-PH-884, MADPH-95-884, IUHET-299",
    doi = "10.1103/PhysRevLett.75.1462",
    journal = "Phys. Rev. Lett.",
    volume = "75",
    pages = "1462--1465",
    year = "1995"
}

@article{bartosik2020performance,
  title={Detector and Physics Performance at a Muon Collider},
  author={Bartosik, N. and others},
  journal={Journal of Instrumentation},
  volume={15},
  number={05},
  pages={P05001},
  year={2020}
}

@article{lucchesi2020detector,
  title={Detector performance studies at a muon collider},
  author={Lucchesi, D. and others},
  journal={PoS EPS-HEP2019},
  pages={118},
  year={2020}
}

@article{collamati2021bib,
  title={Advanced assessment of beam-induced background at a muon collider},
  author={Collamati, F. and Curatolo, C. and Lucchesi, D. and Mereghetti, A. and Mokhov, N. and Palmer, M. and Sala, P.},
  journal={Journal of Instrumentation},
  volume={16},
  number={11},
  pages={P11009},
  year={2021}
}

@article{mokhov2011mdi,
  title={Muon collider interaction region and machine-detector interface design},
  author={Mokhov, N. V. and others},
  journal={arXiv preprint arXiv:1202.3979},
  year={2011}
}

@article{Neuffer:2013wrd,
  author = "Neuffer, D. and Palmer, M. and Alexahin, Y. and Ankenbrandt, C. and Delahaye, J. P.",
    title = "{A Muon Collider as a Higgs Factory}",
    booktitle = "{4th International Particle Accelerator Conference}",
    eprint = "1502.02042",
    archivePrefix = "arXiv",
    primaryClass = "physics.acc-ph",
    reportNumber = "FERMILAB-CONF-13-140-APC, IPAC-2013-TUPFI056",
    year = "2013"
}

@article{costantini2020vectorboson,
  author = "Costantini, Antonio and De Lillo, Federico and Maltoni, Fabio and Mantani, Luca and Mattelaer, Olivier and Ruiz, Richard and Zhao, Xiaoran",
    title = "{Vector boson fusion at multi-TeV muon colliders}",
    eprint = "2005.10289",
    archivePrefix = "arXiv",
    primaryClass = "hep-ph",
    reportNumber = "CP3-20-20, MCNET-20-12, VBSCAN-PUB-03-20",
    doi = "10.1007/JHEP09(2020)080",
    journal = "JHEP",
    volume = "09",
    pages = "080",
    year = "2020"
}

@article{black2024forum,
 author = "Black, K. M. and others",
    title = "{Muon Collider Forum report}",
    eprint = "2209.01318",
    archivePrefix = "arXiv",
    primaryClass = "hep-ex",
    reportNumber = "FERMILAB-FN-1194",
    doi = "10.1088/1748-0221/19/02/T02015",
    journal = "JINST",
    volume = "19",
    number = "02",
    pages = "T02015",
    year = "2024"
}

@article{accettura2024imcc,
  title={Interim report for the International Muon Collider Collaboration},
  author={Accettura, C. and others},
  journal={arXiv preprint arXiv:2407.12450},
  year={2024}
}

@article{es2023ragas,
  title={RAGAS: Automated Evaluation of Retrieval Augmented Generation},
  author={Es, Shahul and James, Jithin and Espinosa-Anke, Luis and Schockaert, Steven},
  journal={arXiv preprint arXiv:2309.15217},
  year={2023}
}

@inproceedings{lin2004rouge,
  title={ROUGE: A Package for Automatic Evaluation of Summaries},
  author={Lin, Chin-Yew},
  booktitle={Text Summarization Branches Out},
  pages={74--81},
  year={2004}
}

@inproceedings{voorhees1999trec,
  title={The TREC-8 Question Answering Track Report},
  author={Voorhees, Ellen M.},
  booktitle={Proceedings of TREC},
  year={1999}
}

@article{jarvelin2002ndcg,
  title={Cumulated gain-based evaluation of IR techniques},
  author={J{\"a}rvelin, Kalervo and Kek{\"a}l{\"a}inen, Jaana},
  journal={ACM Transactions on Information Systems},
  volume={20},
  number={4},
  pages={422--446},
  year={2002}
}

@article{han2020higgs,
  title         = {Electroweak Couplings of the Higgs Boson at a Multi-TeV Muon Collider},
  author        = {Han, Tao and Liu, Da and Low, Ian and Wang, Xing},
  journal       = {Physical Review D},
  volume        = {103},
  pages         = {013002},
  year          = {2021},
  eprint        = {2008.12204},
  archivePrefix = {arXiv},
  primaryClass  = {hep-ph}
}

@article{celada2024higgs,
  title         = {Probing Higgs-muon interactions at a multi-TeV muon collider},
  author        = {Celada, E. and others},
  journal       = {Journal of High Energy Physics},
  volume        = {2024},
  number        = {8},
  pages         = {21},
  year          = {2024}
}

@article{abbott2023anomalous,
  title         = {Anomalous production of massive gauge boson pairs at muon colliders},
  author        = {Abbott, B. and others},
  journal       = {Physical Review D},
  volume        = {108},
  pages         = {093009},
  year          = {2023}
}

@inproceedings{beltagy2019scibert,
  title     = {SciBERT: A Pretrained Language Model for Scientific Text},
  author    = {Beltagy, Iz and Lo, Kyle and Cohan, Arman},
  booktitle = {Proceedings of EMNLP-IJCNLP},
  pages     = {3615--3620},
  year      = {2019}
}

@inproceedings{jain2020s2orc,
  title     = {S2ORC: The Semantic Scholar Open Research Corpus},
  author    = {Lo, Kyle and Wang, Lucy Lu and Neumann, Mark and Kinney, Rodney and Weld, Daniel S.},
  booktitle = {Proceedings of ACL},
  pages     = {4969--4983},
  year      = {2020}
}

@inproceedings{lopez2009grobid,
  title     = {GROBID: Combining Automatic Bibliographic Data Recognition and Term Extraction for Scholarship Publications},
  author    = {Lopez, Patrice},
  booktitle = {Proceedings of ECDL},
  pages     = {473--474},
  year      = {2009}
}

@article{ji2023survey,
  title={Survey of hallucination in natural language generation},
  author={Ji, Ziwei and Lee, Nayeon and Frieske, Rita and Yu, Tiezheng and Su, Dan and Xu, Yan and Ishii, Etsuko and Bang, Yejin and Madotto, Andrea and Fung, Pascale},
  journal={ACM Computing Surveys},
  volume={55},
  number={12},
  pages={1--38},
  year={2023}
}

@article{gendreaudistler2025automatinghighenergyphysics,
      title={Automating High Energy Physics Data Analysis with LLM-Powered Agents}, 
      author={Eli Gendreau-Distler and Joshua Ho and Dongwon Kim and Luc Tomas Le Pottier and Haichen Wang and Chengxi Yang},
      year={2025},
      journal={arXiv preprint arXiv:2512.07785},
      archivePrefix={arXiv},
      primaryClass={physics.data-an},
}

@article{moreno2026aiagentsautonomouslyperform,
      title={AI Agents Can Already Autonomously Perform Experimental High Energy Physics}, 
      author={Eric A. Moreno and Samuel Bright-Thonney and Andrzej Novak and Dolores Garcia and Philip Harris},
      year={2026},
      journal={arXiv preprint arXiv:2603.20179},
      eprint={2603.20179},
      archivePrefix={arXiv},
      primaryClass={hep-ex},
}

@article{gao2024retrievalaugmentedgenerationlargelanguage,
      title={Retrieval-Augmented Generation for Large Language Models: A Survey}, 
      author={Yunfan Gao and Yun Xiong and Xinyu Gao and Kangxiang Jia and Jinliu Pan and Yuxi Bi and Yi Dai and Jiawei Sun and Meng Wang and Haofen Wang},
      year={2024},
      journal={arXiv preprint arXiv:2312.10997},
      archivePrefix={arXiv},
      primaryClass={cs.CL},
}

@article{robertson-2009,
	author = {Robertson, Stephen and Zaragoza, Hugo},
	journal = {Foundations and Trends® in Information Retrieval},
	number = {1-2},
	pages = {1--174},
	title = {{The Probabilistic Relevance Framework: BM25 and beyond}},
	volume = {4},
	year = {2009},
	doi = {10.1561/1500000019},
}

@article{McGreivy:2025rrz,
     author = "McGreivy, James and Delaney, Blaise and Beck, Anja and Williams, Mike",
    title = "{Seeing the Forest Through the Trees: Knowledge Retrieval for Streamlining Particle Physics Analysis}",
    eprint = "2509.06855",
    archivePrefix = "arXiv",
    primaryClass = "hep-ex",
    year = "2025"
}

@inproceedings{Mallampalli:2026hrl,
    author = "Mallampalli, Abhishikth and Dasu, Sridhara",
    title = "{MITRA: An AI Assistant for Knowledge Retrieval in Physics Collaborations}",
    booktitle = "{39th Annual Conference on Neural Information Processing Systems}: {Includes Machine Learning and the Physical Sciences (ML4PS)}",
    eprint = "2603.09800",
    archivePrefix = "arXiv",
    primaryClass = "cs.IR",
    year = "2026"
}

@article{Qian:2021ihf,
    author = "Qian, Sitian and Li, Congqiao and Li, Qiang and Meng, Fanqiang and Xiao, Jie and Yang, Tianyi and Lu, Meng and You, Zhengyun",
    title = "{Searching for heavy leptoquarks at a muon collider}",
    eprint = "2109.01265",
    archivePrefix = "arXiv",
    primaryClass = "hep-ph",
    doi = "10.1007/JHEP12(2021)047",
    journal = "JHEP",
    volume = "12",
    pages = "047",
    year = "2021"
}

@article{Jiang:2023mte,
    author = "Jiang, Ruobing and Yang, Tianyi and Qian, Sitian and Ban, Yong and Li, Jingshu and You, Zhengyun and Li, Qiang",
    title = "{Searching for Majorana neutrinos at a same-sign muon collider}",
    eprint = "2304.04483",
    archivePrefix = "arXiv",
    primaryClass = "hep-ph",
    doi = "10.1103/PhysRevD.109.035020",
    journal = "Phys. Rev. D",
    volume = "109",
    number = "3",
    pages = "035020",
    year = "2024"
}

@article{Jiang:2024wwa,
    author = "Jiang, Ruobing and Jiang, Chuqiao and Ruzi, Alim and Yang, Tianyi and Ban, Yong and Li, Qiang",
    title = "{Searches for multi-Z boson productions and anomalous gauge boson couplings at a muon collider}",
    eprint = "2404.02613",
    archivePrefix = "arXiv",
    primaryClass = "hep-ex",
    doi = "10.1088/1674-1137/ad5661",
    journal = "Chin. Phys. C",
    volume = "48",
    number = "10",
    pages = "103102",
    year = "2024"
}

@inproceedings{cormack2009reciprocal,
  author    = {Gordon V. Cormack and Charles L. A. Clarke and Stefan B{\"u}ttcher},
  title     = {Reciprocal Rank Fusion Outperforms {Condorcet} and
               Individual Rank Learning Methods},
  booktitle = {Proceedings of the 32nd Annual International ACM SIGIR Conference},
  pages     = {758--759},
  year      = {2009},
  publisher = {ACM},
  doi       = {10.1145/1571941.1572114}
}

@article{pradeep2021rrf102,
  author  = {Ronak Pradeep and Rodrigo Nogueira and Jimmy Lin},
  title   = {{RRF102}: Meeting the {TREC-COVID} Challenge with a
             100+ Runs Ensemble},
  journal = {arXiv preprint arXiv:2010.00200},
  year    = {2021}
}

@inproceedings{askeda2024,
  author  = {Ramtin Mesbahi and others},
  title   = {Ask-{EDA}: A Design Assistant Empowered by {LLM},
             Hybrid {RAG} and Abbreviation De-hallucination},
  journal = {arXiv preprint arXiv:2406.06575},
  year    = {2024}
}

@techreport{muoncollider_pbc2020,
  author      = {Delahaye, J.-P. and others},
  title       = {Muon Colliders},
  institution = {CERN},
  note        = {arXiv:1901.06150},
  year        = {2019}
}

@techreport{imcc2022,
  author      = {Aime, C. and others},
  title       = {Muon Collider Physics Summary},
  institution = {International Muon Collider Collaboration},
  note        = {arXiv:2203.07256},
  year        = {2022}
}

\end{document}